\documentclass[a4paper,11pt]{article}
\usepackage{pos}
\usepackage{bbold}
\usepackage{graphicx}
\newcommand{\bea}{\begin{eqnarray}}
\newcommand{\eea}{\end{eqnarray}} 
\newcommand{\beq}{\begin{equation}}
\newcommand{\eeq}{\end{equation}} 

\newcommand{\bx}{\mathbf{x}}

\newcommand{\vx}{\bx}

\newcommand{\tr}{\mbox{Tr}}
\newcommand{\bs}{\bigskip}
\newcommand{\vz}{\mathbf{z}}

\newcommand{\U}{\mathcal{U}}

\renewcommand{\b}{\beta}

\newcommand{\m}{\mu}
\newcommand{\g}{\gamma}

\renewcommand{\d}{\delta}

\newcommand{\oh}{\frac{1}{2}}

\newcommand{\dg}{\dagger}
\newcommand{\non}{\nonumber}

\newcommand{\rf}[1]{(\ref{#1})}
\newcommand{\ra}{\rightarrow}
\newcommand{\pa}{\partial}
\newcommand{\pbar}{\overline{\psi}}
\renewcommand{\th}{\theta}
\newcommand{\Sc}{S$_c$}
\title{Excitations of gauge bosons, and other aspects of gauge Higgs theories}

\author*[a]{Jeff Greensite}

\affiliation[a]{Physics and Astronomy Dept., San Francisco State University,\\
  1600 Holloway Avenue, San Francisco CA, USA}

\emailAdd{jgreensite@gmail.com}

\abstract{I review the qualitative physical distinction between the Higgs and confinement phases of a gauge Higgs theory,
and the non-local order parameter introduced by Matsuyama and myself which identifies the two phases.  I then present some new results suggestive of a possible excitation spectrum of vector bosons in the electroweak sector of the
Standard Model.}

\FullConference{The XVIth Quark Confinement and the Hadron Spectrum Conference (QCHSC24)\\
 19-24 August, 2024\\
 Cairns Convention Centre, Cairns, Queensland, Australia\\}

\begin{document}

\maketitle

\section{Introduction}

   Not all gauge transformations are created equal.  There is a special subset, namely the global center subgroup of the gauge group (``GCS'' hereafter), which leaves the gauge field invariant, transforming only matter fields.  A subgroup of this kind is strongly reminiscent of a global internal symmetry of the matter fields, and like any global symmetry (and unlike a local symmetry) it may break spontaneously.  The challenge is to find a locally gauge-invariant order parameter for the breaking, and to understand the physical distinction between broken and unbroken phases.  In ref.\  \cite{Greensite:2020nhg} Matsuyama and I claim to have constructed this order parameter (it is related to the Edwards-Anderson order parameter for spin glasses \cite{Edward_Anderson}), which identifies the Higgs phase of a gauge Higgs theory, and to understand the transition between the confinement and Higgs phases as a transition between different types of confinement.  In this talk I will briefly review these ideas, which are now several years old, and then turn to some new work concerning possible new excitations of gauge bosons in the electroweak sector of the Standard Model.  That new work is described in more detail in a recent preprint \cite{Greensite:2025fdq}.
 
 \section{Confinement to Higgs transition and the global center subgroup}
 
   Let us begin with the distinction between charged and uncharged states in an infinite volume.  This is a matter of convenience since,  strictly speaking, there are no charged states in a finite periodic volume.  The question is whether a charged state can be distinguished from a color neutral state by its transformation properties under the gauge group.  At first sight, a state transforming in this way would seem to violate the physical state condition, i.e.\ the Gauss Law.  But in fact there is an important loophole:  a charged state {\it can} transform under the GCS with violating the physical state condition. 
The simplest example is the state consisting of an electrically charged static particle coupled to the quantized Maxwell field.  The ground state of such a system is \cite{Dirac:1955uv}
\beq
  \Psi_{\text{chrg}} =  \pbar(x) \rho(x;A) \Psi_0 \ ,
\eeq
where
\beq  
\rho(x;A) =  \exp\left[-i {e\over 4\pi} \int d^3z ~ A_i(\vz) {\pa \over \pa z_i}  {1\over |\vx-\vz|}  \right] \ ,
 \label{rho}
 \eeq
 with $\Psi_0$ is the ground state and $\pbar$ a static fermion.  It is easy to check that this state satisfies the Gauss Law, and is invariant under local gauge transformations.   But the GCS of the U(1) gauge group is all transformations $g(x) = e^{i\th}$, and the gauge field does not transform under those transformations, while the matter field does.  Therefore, under a GCS transformation $\Psi_{\text{chrg}} \ra e^{-i\th} \Psi_{\text{chrg}}$.  This is in
contrast to an electrically neutral physical state in, e.g., the abelian Higgs theory, such as $\Psi_{neutral}=\pbar(x) \phi(x)\Psi_0$, which is invariant under all gauge transformations including the GCS.  These considerations generalize readily to a non-abelian theory.    Let us consider a gauge Higgs theory with the scalar field $\phi$ in the fundamental representation of the gauge group, and a static charge created by $\pbar$ in the conjugate representation.  We also consider ``pseudomatter'' fields  $\xi(x;U)$ which are functionals of the lattice gauge field $U$ alone, and which transform like a matter field in the fundamental representation {\it except} that that they are invariant GCS transformations.  Then
\beq
\Psi_{neutral} = \pbar(x) \phi(x) \Psi_0 ~~~~~~~\mbox{and} ~~~~~~~~ \Psi_{chrg} = \pbar(x) \xi(x;U) \Psi_0
\eeq
are examples, respectively, of a color neutral state invariant under all gauge transformations, and a charged state which transforms under the GCS.  For example,
in an SU(N) gauge theory, the GCS consists of global gauge transformations $g(x)=z\mathbb{1}$, where $z\in Z_N$, and $\Psi_{chrg} \ra z \Psi_{chrg}$.
The functional $\rho(x;A)$ in \rf{rho} is one example of a pseudomatter field in U(1) gauge theory.  Gauge transformations to a physical (e.g.\ Couloumb) gauge are another example, and a third, which we use extensively below, are the eigenstates $\xi_n(x)$
\beq
D_{xy} \xi_n(y;U) = \lambda_n \xi_n(x;U)
\eeq
 of the covariant lattice Laplacian
 \beq
 D_{xy} = \sum_{k=1}^3 \left[\mathbb{1} \d^{ab} \d_{xy} - U_k(x) \d_{y,x+\hat{k}}  
- U_k^{\dg}(x-\hat{k}) \d_{y,x-\hat{k}}   \right] \ .
\label{Laplacian}
\eeq

The energy of a charged state in an infinite volume will of course depend on the choice of $\xi$, but what is the 
minimum energy $E_{min}$ of such states?   There are three possibilities:
\begin{enumerate}
\item $E_{min}=\infty$ for any choice of $\xi(x;U)$, and the GCS symmetry is unbroken.  We call this property ``separation of charge'' (or \Sc) confinement, and a theory with this property,
including gauge Higgs theories, is in the confinement phase.  There are no isolated color charges; all color charges exist only in color neutral combinations, such as fermions with Higgs fields and fermions with antifermions, and all particle states are invariant under the GCS.
\item $E_{min}$ is finite for some choices of $\xi$, and the GCS symmetry is unbroken.  Then the theory is in the Coulomb phase, and there can be
finite energy charged states transforming under the GCS.
\item $E_{min}$ is finite, and the GCS symmetry is spontaneously broken.  In this case the sharp distinction between charged and uncharged
states breaks down, since, in the absence of unbroken GCS symmetry the overlap $\langle \Psi_{neutral}|\Psi_{chrg}\rangle$ is not necessarily zero.
This is the Higgs phase.  The fact that a spontaneously broken GCS implies the absence of \Sc \ confinement was shown in  \cite{Greensite:2020nhg}.
\end{enumerate}

\begin{figure}[htb]
 \centerline{\includegraphics[scale=0.3]{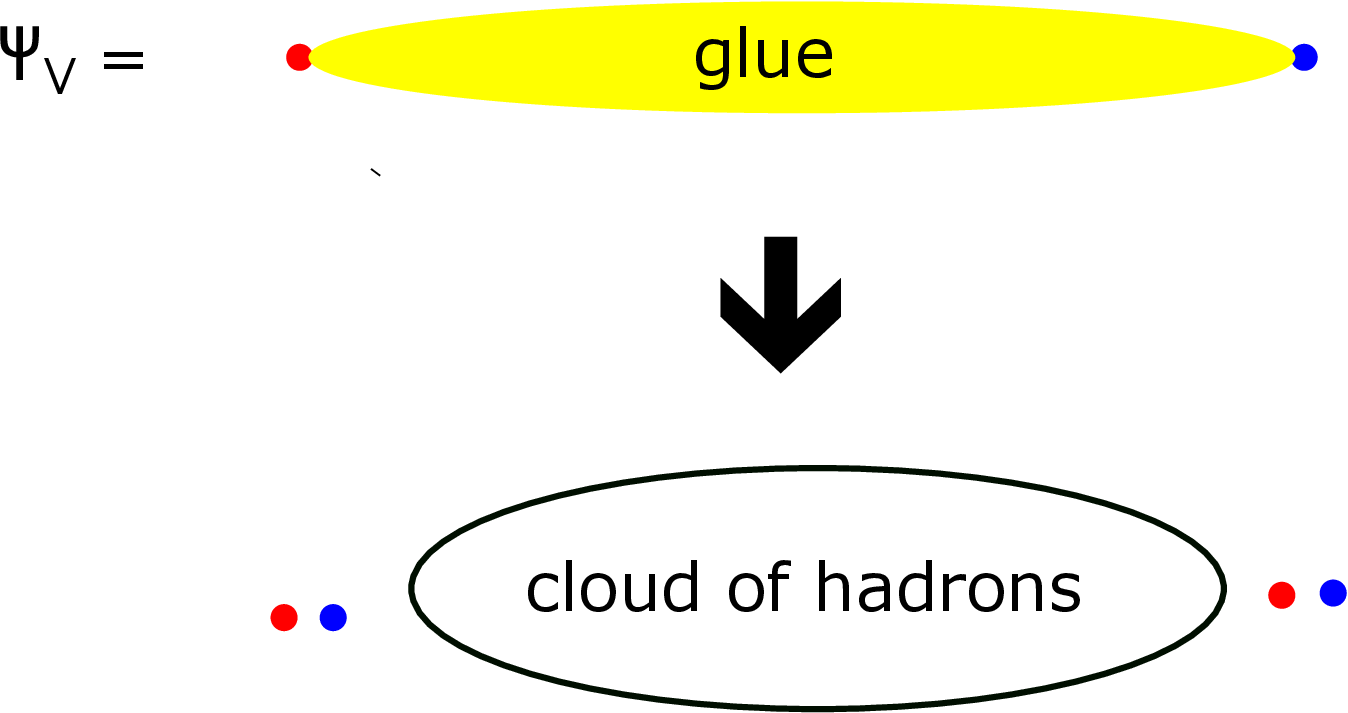}}
 \caption{We are interested in the minimal energy of states like $\Psi_V$, containing isolated color charges connected
 by the gauge field, as color charge separation becomes large. The question is not whether states of this kind decay rapidly; in QCD
 they obviously do.  But such states exist in the Hilbert space, and it makes sense to ask whether their energies increase without limit as separation increases. }
 \label{psiV}
\end{figure}

Thus the transition from the confinement to the Higgs phase is associated with the loss of the \Sc \ confinement property.  Physically this would correspond to linear Regge trajectories in the confined phase, and the absence of such trajectories in the Higgs phase.  This is understood as follows.
Although in a confinement phase such as QCD it is in practice impossible to separate quarks and antiquarks far apart, due to string breaking, it should be recognized that there do exist in the Hilbert space physical states of this kind, of the form
\beq
\Psi_{\xi \xi^\dg,xy} = \pbar(x) \xi(x;U) \xi^\dg(y;U) \psi(y) \Psi_0 \ ,
\eeq
or more generally
\beq \Psi_{V,xy} = \pbar(x) V(x,y;U) \psi(y) \Psi_0  \ ,
\eeq
where $V(x,y,U) \ra g(x) V(x,y;U) g^\dg(y)$ under a local gauge transformation, and is a functional of the gauge field only.  An isolated charge at $x$ would correspond to the limit $y\ra \infty$.  For large separations $R = |x-y|$ such a state (indicated by the top drawing in Fig.\ \ref{psiV}) is of course very unstable in the confinement phase and in, e.g., QCD, would immediately decay via string breaking into  conventional hadrons.   Nevertheless we can ask how the energy $E_V(R)$ of states of the form $\Psi_V$ behaves as $R$ increases, because this property probes the IR structure of the ground state.  If $E_V(R)\ra \infty$ as $R\ra \infty$ for {\it any} choice of $V$, then the theory has the \Sc \ property and is in the confined phase, regardless of whether the theory contains matter in the fundamental representation.  And in fact,
resonances which in QCD lie on linear Regge trajectories are states of this sort, color electric flux is collimated into tubes, and the limitless growth of energy with $R$ is only foiled by string breaking.  This is not the case in the Higgs phase; there are no metastable flux tubes, and no string to break.

    If the transition to the Higgs phase corresponds to the spontaneous breaking of GCS symmetry, the next question is the order parameter.  It has been known since the work of Osterwalder and Seiler \cite{Osterwalder:1977pc}, later elaborated in \cite{Fradkin:1978dv,Banks:1979fi}, that the Higgs and confinement phases are not entirely separated by a thermodynamic transition.  This fact should not be taken to mean that there is no transition between
physically distinct phases in the absence of non-analyticity in the free energy; the Kertesz line in Ising models is a well-known counter-example \cite{Kertesz}.  It does mean, however, that we cannot expect to find a local order parameter for the transition.  The construction of the order parameter is roughly analogous to the Edwards-Anderson order parameter in spin glasses \cite{Edward_Anderson}, with the ``spins''  corresponding to the scalar field on sites, and the random couplings to the link  variables.  Without elaborating on the analogy here, we simply describe the construction.  Let 
$\U_k(x)=U_k(x,t=0)$ be the gauge field on the $t=0$ time slice, with $\Psi_0[\U,\phi]$ the ground state of a gauge Higgs theory.  Then define
\bea                                
           \overline{\phi}(x;\U) &=& \int D\phi ~ \phi(x) \Psi_0^2(\phi,\U) \label{overline} \non \\ 
            \Phi(\U) &=& {1\over V} \sum_x  | \overline{\phi}(x;\U) | \label{max}  \non \\
            \langle \Phi \rangle &=& {1\over Z} \int DU D\phi ~ \Phi(\U) e^{-S}
\eea
In words: $\overline{\phi}(x;\U)$ is the expectation value of $\phi(x)$ on some time slice, with the gauge field on that
time slice held fixed.  With fixed $\U$, there is still a remaining symmetry of the theory, namely the GCS.  If 
$\overline{\phi}(x;\U)$ is non-zero (but in general different) at each $x$, then the GCS is spontaneously broken, and if
this breaking is generally true for backgrounds $\U$ generated by the usual Boltzman distribution, then we conclude that the GCS is spontaneously broken in the gauge Higgs theory.  Note that this definition does not require a gauge choice on the time slice.  In fact, $\overline{\phi}(x;g\circ \U) \ra g\overline{\phi}(x;\U)$ transforms covariantly, and 
$\Phi(\U)$ is gauge
invariant.  The expectation value $\langle \Phi \rangle$ of this gauge invariant observable can be computed by a
lattice Monte Carlo procedure, and $\langle \Phi \rangle > 0$ implies the loss of \Sc ~ confinement, as discussed in  \cite{Greensite:2020nhg}.  This allows us to find numerically the transition line separating the Higgs and confinement phases in, e.g., SU(2) gauge Higgs theory. See also \cite{Alles:2024qha} for a recent application in $Z_2$ lattice gauge Higgs theory in three dimensions, and \cite{Ward:2021qqh} for the transition line in SU(2) gauge Higgs theory in five Euclidean dimensions.

\section{A spectrum of gauge bosons in the electroweak theory?}

   I will now turn to the second topic of these proceedings.  We begin from the fact that any elementary particle in a gauge Higgs theory is in some sense a composite object, consisting of, e.g., a charged object surrounded by gauge and Higgs fields.  It is generally true in quantum mechanics that composite objects have a discrete spectrum, and this leads to the conjecture that what we call ``elementary'' particles may themselves have a discrete spectrum of excitations.
We would like to investigate this conjecture numerically, for vector bosons in the quenched electroweak sector of the
Standard Model.

The idea is to diagonalize the transfer matrix in a certain subspace of the Hilbert Space of the quenched electroweak sector which contains, in particular, photon and Z boson states.  The question is whether there are any other particle  states in the spectrum, which are neither multiparticle states, nor states of known particles at finite momenta.  The
lattice action is \cite{Zubkov:2008gi,Shrock:1985ur}
\bea
 & &S  = -\beta \sum_{plaq} \left[\oh \tr[UUU^\dg U^\dg] + {1\over \tan^2(\th_W)} \text{Re}[VVV^\dg V^\dg] \right] \non \\
      & & \ \  -2\sum_{x,\m} \text{Re}[\phi^\dg(x) U_\m(x) V_\m(x) \phi(x+\hat{\m})]  
       - \sum_x \{ -(\g-8) \phi^\dg(x)\phi(x)  + \l (\phi^\dg(x)\phi(x))^2 \} \ ,
 \eea
 with SU(2) gauge field $U_\m(x)$, U(1) gauge field ${V_\m(x) = e^{i\th_\m(x)}}$, and Higgs field $\phi(x)$, with $\th_W$ the Weinberg angle. Phenomenology gives the tree level values
\beq
 \sin^2\th_W = 0.231 ~~,~~ \beta = 10.1 ~~,~~ \lambda = 0.13 \ .
 \eeq
 while $\g$, which corresponds to a dimensionful quantity in the continuum Higgs potential, depends on the lattice spacing.  Let  $\tilde{U}=UV$ be the SU(2)$\times$ U(1) gauge field,  and $D^{ab}_{xy}[\tilde{U}]$ the covariant lattice Laplacian.  Denote $\zeta_1(x) = \phi(x)$ the Higgs field, and $\{ \zeta_n(x) , n=2,...,n_{ev}+1 \}$,  are the first $n_{ev}$  eigenstates of the covariant Laplacian.    In this calculation we use a maximum of $n_{ev}=32$.  We create
a subspace of excited states spanned by (non-orthogonal) states $\{ |\Phi^n_\m \rangle, n=1,2,...,11 \}$ \bs
\bea
    \eta(\vx) e^{i{\cal A}^n_\m(\vx)} &=& \zeta_n^\dg(x) \tilde{U}_\m(\vx,t) \zeta_n(\vx+\hat{\m})  \non \\
     A^n_\m(\vx) &=& \sin({\cal A}^i_\m(\vx)) \non \\
     Q_\m^n &=& {1\over L^3} \sum_\vx A^n_\m(\vx) \non \\
     |\Phi_\m^n\rangle &=& Q^n_\m |\Psi_0\rangle \ .
\label{Q}
\eea  
In order to diagonalize the transfer matrix in the subspace, we define
\bea
            O_{ab} &=&  \langle\Phi_\m^a|\Phi_\m^b\rangle = \langle Q^{a\dg}_\m(t) Q^b_\m(t) \rangle \non \\
            T_{ab}  &=& \langle \Phi_\m^a|e^{-(H-E_0)}|\Phi_\m^b\rangle =  \langle Q^{a\dg}_\m(t+1)  Q^b_\m(t) \rangle \ ,
 \eea
 where $e^{-(H-E_0)}$ is the transfer matrix with the vacuum energy removed.  The right hand sides of these two
 expressions are computed by lattice Monte Carlo.  We then solve numerically the generalized eigenvalue problem
  \beq
            [T] \vec{\upsilon}^n = \lambda_n [O] \vec{\upsilon}^n \ ,
 \eeq 
 and then
 \beq
              |\Psi^n_\m\rangle = \sum_a \upsilon^n_a |\Phi_\m^a\rangle 
 \eeq
 are the eigenstates of the transfer matrix in this subspace, with masses
 \beq
 M_n = -\log(\lambda_n) \ .
 \eeq
 We are free to vary the number $n_{ev} \le 32$ included in the calculation, to see whether $M_n$ reaches a plateau with increasing $n_{ev}$.
 
\begin{figure}[htb]
 \centerline{\includegraphics[scale=0.6]{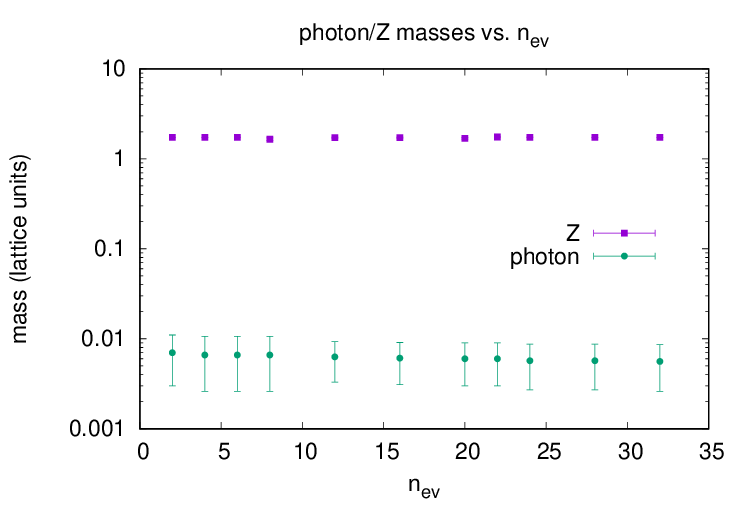}}
 \caption{Photon and Z mass in lattice units vs.\ $n_{ev}$ at $\g=4$ on a $16^3\times 72$ lattice volume.}
 \label{pZ}
\end{figure}

    The state which we identify as the Z boson appears already at $n_{ev}=0$.   At $\g=4$ its mass $m_Z$ in  lattice units is $1.735(2)$, and a state of this same mass can be identified among the excited levels as $n_{ev}$ increases, as shown in Fig.\ \ref{pZ} for $\g=4$.  The $Z$ mass at tree-level is
\beq
   m^{tree}_Z =   {1\over \cos \th_W} \sqrt{\g \over \lambda \b} \ ,
\label{mZ}
\eeq
and at $\g=4$ the ratio $m_Z/m^{tree}_Z$ is 0.87, so these values are fairly close.  We therefore tentatively identify states of this mass, appearing already at $n_{ev}=0$, as the Z boson.

   At $n_{ev}=2$ a very light mass state appears, and this lowest mass state remains very light as $n_{ev}$ increases, as also shown in Fig.\ \ref{pZ}.  This mass is lattice size dependent, and drops by about a factor of three in going from
a $12^3 \times 72$ lattice volume to a $16^3 \times 72$ volume.  Therefore there is some reason to believe that
the finite mass is a finite size effect, and will drop to zero in the infinite volume limit.  We identify this state as the photon.

\begin{figure}[htb]
  \centerline{\includegraphics[scale=0.6]{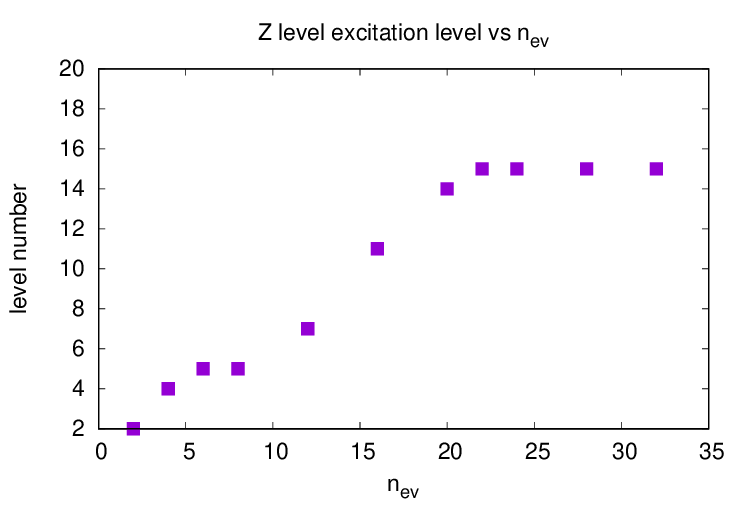}}
 \caption{Excitation number of the Z boson vs.\ $n_{ev}$, showing how new excitations appear in the spectrum between
 the photon and the $Z$ with increasing the number of pseudomatter fields.} 
 \label{Zlevel}
\end{figure}

   As $n_{ev}$ increases, more states appear in the spectrum between the photon and the state we identify, by its mass in lattice units, as the Z.  
Fig.\ \ref{Zlevel} displays the excitation number of the state identified as the Z versus $n_{ev}$ ($\g=4$, and $16^3\times 72$ volume), which increases with $n_{ev}$ as more states enter the spectrum between the photon and the $Z$. At $n_{ev}=22$ and above, the level number of the $Z$ stabilizes at level 15.

 \begin{figure}[htb]
  \centerline{\includegraphics[scale=0.6]{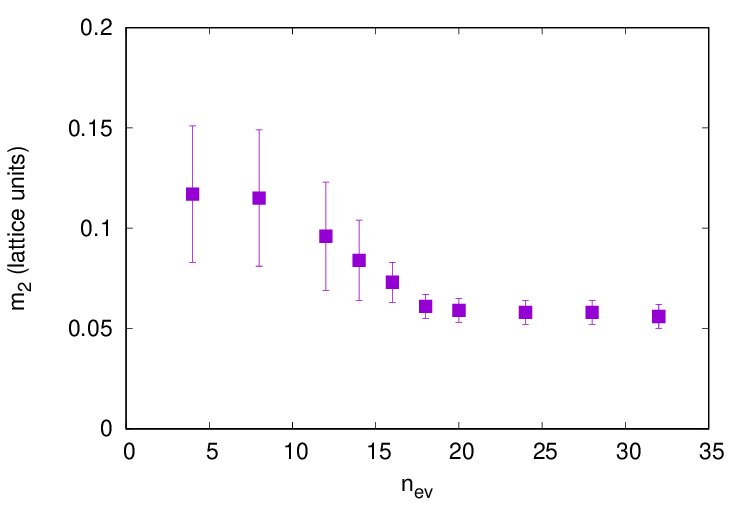}}
 \caption{Mass (in lattice units) of the first excitation above the photon state, vs.\ the number $n_{ev}$ of pseudomatter states spanning the truncated
 space of states.  Data is for $\g=4$, and we have convergence by $n_{ev}=20$.} 
 \label{m2}
\end{figure}  

 \begin{figure}[htb]
  \centerline{\includegraphics[scale=0.6]{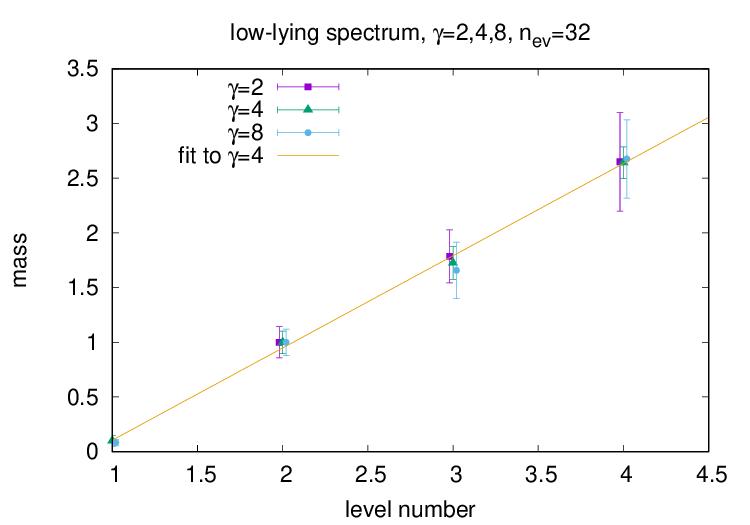}}
 \caption{Energies of the first four energy levels in the spectrum obtained from the generalized eigenvalue calculation (with $n_{ev}=32$).
 Results are shown for $\g=2,4,8$, in units where the mass of the second level is $m_2=1$.  The lattice volume is $16^3\times 72$, and
 the straight line is a fit to the $\g=4$ data points. Data points are slightly displaced horizontally for visibility.}
 \label{gall_4}
\end{figure}   

    Figure \ref{m2} displays the mass $m_2$ of the first excitation above the photon state in lattice units vs.\ $n_{ev}$, which converges at $n_{ev}=20$.   This cannot be identified as two photons of opposite momenta, because for a lattice of spatial extension 16 this would be an energy of at least ${4\pi\over 16} = 0.785$ in lattice units.   For comparison, the mass of $m_2$ in lattice units at $\g=4$ is 0.056, and the ``two photon'' interpretation is untenable. In Fig.\ \ref{gall_4} we show the first four energy levels in the spectrum at $\g=2,4,8$, in units of the mass of the $n=2$ energy level $m_2$.  This is the level just above the photon state, and we set the scale so that $m_2=1$.  These first four excitations are fit fairly accurately by a straight line fit to the $\g=4$ data, with a slope not far from one and intercept not far from zero.  The interpretation is that the third and fourth excitations correspond to two and three particles at zero momentum, each with the mass $m_2$ of the excited state above the photon.
The next question is what is that mass, in physical units, and here uncertainties enter the picture.  The problem is that
if we set units such that $m_2=1$, the masses of the higher excitations disagree at different $\g$.  Taking the level 15
excitation as the Z boson, and using the Z mass to set the scale, we find the results for $m_2$ shown in Table I.

\begin{table}[t!]
\begin{center}
\begin{tabular}{|c|c|c|} \hline
         $m_2$ (GeV) &  $ \g $ & lattice volume  \\
\hline
        3.6(5) &          2  &  $16^3 \times 72$  \\
        3.0(3) &          4  &  $16^3 \times 72$  \\
        3.6(4) &          8  &  $16^3 \times 72$ \\
        4.0(4) &          2  &  $12^3 \times 72 $ \\
        3.5(3) &          4  &  $12^3 \times 72$  \\
        3.3(3) &          8  &  $12^3 \times 72$  \\       
\hline
\end{tabular}
\caption{Mass $m_2$ of the lightest state above the photon, computed at three different $\g$ values and two
lattice volumes.  In each case the physical scale is set by identifying the level 15 state with the Z boson (see text).} 
\label{tab1}
\end{center}
\end{table}

\newpage

All that can be said here is that the mass of the lowest excitation above the photon seems to be somewhere in the
range of 3-4 GeV.  This is not a very helpful prediction from the experimental point of view, given the uncertainty in mass, and complete ignorance of lifetime.  Its significance at this stage is mainly as a strong suggestion.  There seems no reason, in principle, that there shouldn't be a spectrum of vector boson excitations beyond the photon and the $Z$, and here we have numerical evidence in support of that conjecture.  But much more work is required to improve the robustness and reliability of this calculation, as discussed in \cite{Greensite:2025fdq}.  Perhaps there will be more to report along those lines  at the next QCHS meeting.
    
 \acknowledgments{This research is supported by the U.S.\ Department of Energy under Grant No.\ DE-SC0013682.}    
 
\bibliography{sym3.bib}
 
\end{document}